\documentclass{ieeeaccess}
\usepackage{subfigure}
\usepackage{caption}
\usepackage{url}
\usepackage{algorithm}
\usepackage{algpseudocode}
\usepackage{booktabs}

\usepackage{cite}
\usepackage{amsmath,amssymb,amsfonts}
\usepackage{graphicx}
\usepackage{textcomp}
\begin{document}
	\history{Pre-print version}
	\doi{10.1109/ACCESS.2019.2953043}
	
	\title{Exploiting Information-centric Networking to Federate Spatial Databases}
	
	\author{\uppercase{{Andrea Detti}}, 
		\uppercase{Giulio Rossi and Nicola Blefari Melazzi}}
	\address{CNIT/Electronic Engineering Department, University of Rome Tor Vergata, Italy}
	\markboth
	{A. Detti \headeretal: Exploiting Information-centric Networking to Federate Spatial Databases}
	{A. Detti \headeretal: Exploiting Information-centric Networking to Federate Spatial Databases}
	
	\corresp{Corresponding author: Andrea Detti (e-mail: andrea.detti@uniroma2.it).}
	
	\begin{abstract}
		This paper explores the methodologies, challenges, and expected advantages related to the use of the information-centric network (ICN) technology for federating spatial databases. ICN services allow simplifying the design of federation procedures, improving their performance, and providing so-called data-centric security. In this work, we present an architecture that is able to federate spatial databases and evaluate its performance using a real data set coming from OpenStreetMap within a heterogeneous federation formed by MongoDB and CouchBase spatial database systems.
	\end{abstract}
	

	\titlepgskip=-15pt
	
	\maketitle
	
	\section{Introduction} 
	\PARstart{S}{patial} databases play a central role in modern information management systems, managing large volumes of \textit{spatial} objects \cite{SDBintroduction}, which are characterized by a georeferenced geometry (point, polygon, etc.) and a set of properties. Spatial databases supporting spatial functionality are used for many applications, including geographic information system (GIS), navigation software, journey planners, Internet of Things (IoT), etc.
	A typical spatial query searches for objects intersecting an area, or close to a given point, or contained in a given area. Many commercial products exist belonging both to the SQL and NoSQL database families.
	
	A database system (DBS) is normally owned by a single entity that assures the reliability of the offered services and the stored information. Many DBSs also support a \textit{distributed} deployment, usually in a cloud, where a cluster of database nodes forms a common storing space handled by a central database management system (DBMS).
	
\begin{figure}[t] 
		\centering 
		\includegraphics[scale=0.5]{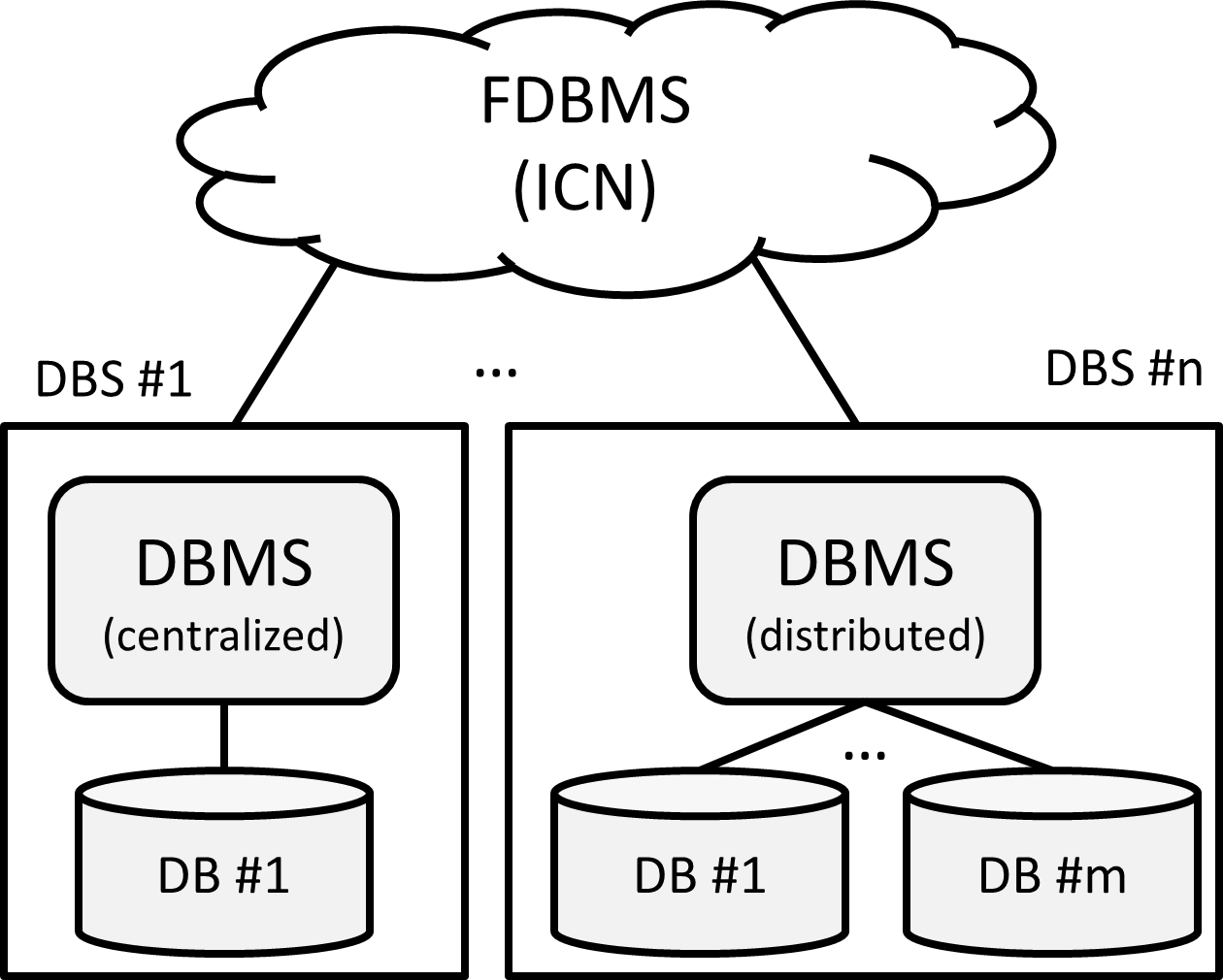} 
		\caption{Federated database system (FDBS)} 
		\label{f:fdbms} 
	\end{figure}

	Instead, a \textit{federated} database system is a collection of cooperating but autonomous component DBSs \cite{sheth1990federated} transparently integrated through a common access interface, as shown in Fig. \ref{f:fdbms}. 
	A database federation can foster the development of services needing to access data stored in separate DBSs and which cannot be merged in a single DB. For instance, organizations managing different information sources are often autonomous and willing to share their data only if they retain control of such data.
	
	A contemporary use case for a federated database comes out from the request of the European Commission to member states to deploy national access points exposing national transport information (traffic status, train schedules, etc.) to foster the development of cross-border multimodal transport services. The federation of these national databases and their capability to appear as a unique database (DB) would simplify the development of cross-border journey planners offering a single point of access to the whole data-set.
	Another use case concerns smart city IoT applications for which the smart behavior of an application often arises out of the integration of cross-domain (health, transport, security, etc.) information stored in different databases. Accordingly, the ETSI ISG CIM group is now discussing how to integrate cross-domain context/IoT information; the federation of different DBSs is a possible solution to such issue. 
	
	The federation of DBSs poses many challenges, including the following:  
	
	\begin{itemize} 
		\item Data heterogeneity - information can be stored with different structures, such as different tags to indicate the same concept, different formats, or different semantics. 
		\item System heterogeneity - the possibility that federated databases have different capabilities, query languages, etc. 
		\item Efficient and secure wide-area communications - the need to design network services to have fast and secure database operations (e.g., queries)
		
	\end{itemize} 
	
	In this paper, we focus on the third challenge described above and propose a solution based on the use of an information-centric network (ICN) \cite{jacobson2009networking} interconnecting the different database sites.
	
	An ICN is a new network layer designed to provide users with \textit{named objects} rather than end-to-end connections. A named object is a bundle of data with a limited size of a few kB, uniquely identified by a hierarchical name.
	To some extent, ICN services resemble those of a content delivery network, but with a finer, packet-level granularity. 
	
	We propose to exploit typical ICN services, such as routing by name, in-network caching, and multicast, to efficiently solve geographical queries and to support a global indexing scheme that shortens query latency and reduce DBS load. We also leverage ICN's data-centric security to ensure provenance and the validity of data and signaling.
	Besides, we implemented our proposed ICN solution and tested it by setting up a federation whose sites use heterogeneous spatial databases with spatial features, namely, the NoSQL MongoDB and CouchBase. The proposed architecture, however, can include other kinds of NoSQL or SQL spatial databases (e.g., PostGIS) if they support the storage of spatial data in GeoJSON data format.

	To the best of our knowledge, this is the first work to propose the use of ICN services for the \textit{federation} of heterogeneous spatial databases. Its main contributions are  
	
	\begin{itemize}     
		\item a methodology for federating spatial databases using the services of an information-centric network,      
		\item a global indexing strategy based on a greedy adaptive tessellation algorithm that enables query routing,     
		\item a real implementation based on two popular databases currently used in production systems, and     
		\item a comprehensive performance evaluation considering heterogeneous databases. 
	\end{itemize}

	   \section{Related concepts and works} 
	\label{s:rw} 
	\begin{figure*}[t] 
		\centering 
		\includegraphics[scale=0.6]{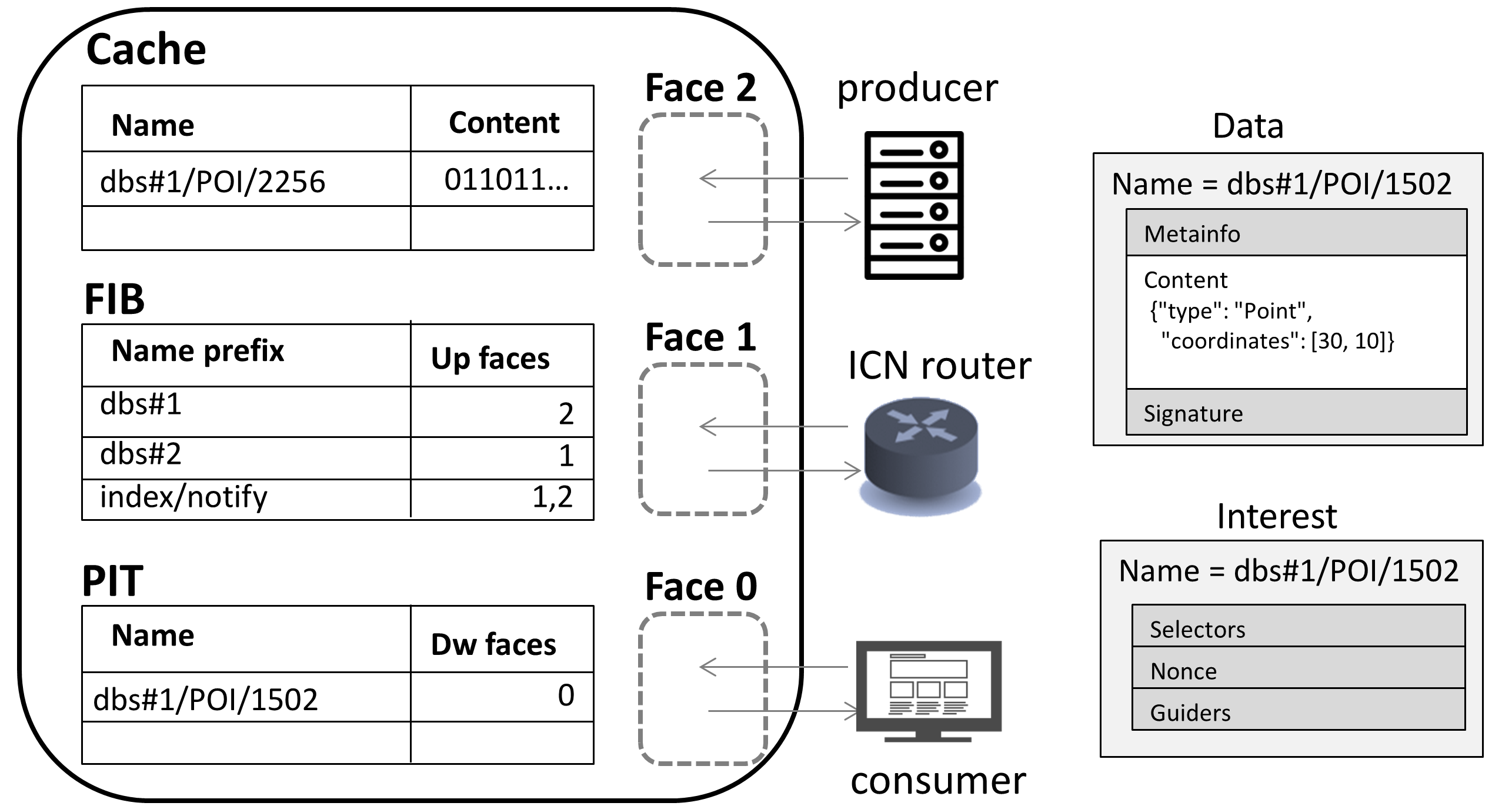} 
		\caption{ICN forwarding engine model and packets (\cite{detti2017application})} 
		\label{f:icn-node} 
	\end{figure*} 
	
	\subsection {Information-centric networks} 
	\label{r:ICN}
	An ICN is a communication architecture providing users with data items rather than end-to-end communication pipes \cite{jacobson2009networking}\cite{MORELLI2017232}. The network addresses are hierarchical names (e.g. \texttt{dbs\#1/poi/1502}) that do not identify end-hosts but data items.
	
	A data item and its unique name form the so-called \textit{named object}. A named object is a small data unit (e.g., 4 KB long) and may contain an entire content (e.g., a document, a video, etc.) or a chunk of it. The names used for addressing the chunks of the same content have a common prefix (e.g., \texttt{dbs\#1/poi/1502}), followed by a sequence number identifier (e.g., \texttt{s1, s2, s3, etc.}).
	
	An ICN is formed by nodes logically classified as consumers, producers, and routers. Consumers pull named objects provided by producers, possibly going through intermediate routers. The consumer-to-producer path is labeled as upstream and the reverse path as downstream. 
	
	Any node uses the forwarding engine shown in Figure \ref{f:icn-node} and is connected to other nodes through channels called \textit{faces}, which can be based on different transport technologies such as Ethernet, TCP/IP sockets, etc.
	
	The data units exchanged in ICN are \textit{Interest packets} and \textit{Data packets}. To download a named object, a consumer issues an Interest packet, which contains the object name and is forwarded to the producer. The forwarding process is referred to as routing by name since the upstream face is selected through a name-based prefix matching based on a \textit{forwarding information base} (FIB) containing name prefixes, such as \texttt{dbs\#1} in Fig. \ref{f:icn-node}.
	The FIB is usually configured by routing protocols, which announce name prefixes rather than IP subnetworks \cite{hoque2013nisr}. During the Interest forwarding process, the node temporarily keeps track of the forwarded Interest packets in a pending information table (PIT), which stores the name of the requested object and the identifier of the face from which the Interest came from (downstream face).
	
	When an Interest reaches a node (producer or an intermediate router) having the requested named object, the node sends back the object within a Data packet whose header includes the object name. The Data packet is forwarded downstream to the consumer by \textit{consuming} (i.e., deleting) the information previously left in the PITs like bread crumbs. 
	
	Each forwarding engine can cache the forwarded Data packet to serve subsequent requests of the same object (\textit{in-network caching}) \cite{ABDULLAHI201548}. Usually, data freshness is loosely controlled by an expiry approach. Any Data packet includes a meta info field reporting the freshness period specified by the producer, which indicates how long the data can be stored in the network cache. 
	
	The forwarding engine also supports \textit{multicast distribution} for both Interest and Data packets. Interest multicasting takes place when there are more upstream faces for a given prefix (e.g., \texttt{index/notify} in Fig. \ref{f:icn-node}), and the incoming Interest is forwarded to all of them. Data multicasting is implemented as follows: when a node receives multiple Interests for the same object, the engine forwards only the first packet and discards the subsequent ones, appending the identifier of the arrival downstream faces in the PIT; then when the requested Data packet arrives, the node forwards a copy of it to each downstream face contained in the PIT.
	
	ICN security is built on the notion of \textit{data-centric} security; the content itself is made secure rather than the connections over which it travels. The ICN security framework provides each entity with a private key and an ICN digital certificate signed by a trust anchor and uniquely identified by a name called \textit{key locator} \cite{ndntrust}. Each Data packet is digitally signed by the content owner and includes the key locator of the digital certificate to be used for signature verification. For access control purposes, Interest packets can be signed too.

	\subsection {Spatial databases} 
	Database systems may be based either on a relational model or a non-relational model also called NoSQL.
	
	Relational databases, such as Oracle MySQL, PostgreSQL/PostGIS, Microsoft SQL, etc., allow complex operations by using a standardized language, that is, the SQL one. Even though every SQL database has its characteristics, to migrate from one engine to another is not so difficult, thus simplifying \textit{database integration}, that is, the merging of different bases in a single one. Many SQL databases support spatial data and related functionality such as range and proximity queries, spatial indexing, etc. 
	
	SQL databases are easy to use when data items are highly structured. On the other side, NoSQL databases have emerged for simplifying the management of heterogeneous data. Indeed, most of NoSQL databases are based on the \textit{document-oriented}, where every document has a unique object identifier, is formed of an arbitrary structure of key/value couples (schema-less), and is usually represented as a JSON object or GeoJSON in case of spatial objects.
	The storage space of a NoSQL database is logically organized in \textit{data-sets}, that is, groups of related objects, such as a MongoDB/DocumentDB collection, a Cassandra column family, or a CouchBase bucket. To some extent, a data-set resembles a SQL table but without a schema. The storage space of a NoSQL can be physically partitioned over different servers (sharding) as a function of the system workload.
	This feature, known as horizontal scalability, fits well with cloud environments, where databases are usually deployed. MongoDB, CouchBase, Cassandra, and DocumentDB are examples of popular NoSQL databases, with the first two offering also spatial features. 
	Different from SQL databases, the NoSQL ones have proprietary interfaces; thus, the change of a database usually requires changing the application too. Moreover, the functionality offered by current NoSQL databases is usually fewer than the one offered by a SQL one, both because SQL databases are much more mature and because SQL data are structured.
	
	An inalienable characteristic of a modern database is the data indexing. An index is an internal data structure where the references of stored data are sorted according to a given field (column) of the data named index key. On one hand, an index consumes storage spaces and computational resources to be sorted; on the other hand, it drastically accelerates queries related to the indexed data fields, especially when the database size grows. For this reason, the indexes should be created only for those fields that are expected to be frequently used in queries \footnote{SQL databases simplify this choice, providing statistics of executed statements/queries.}.
	Spatial databases have specialized index structures that improve the speed of spatial operations. In general, a spatial indexing method partitions the geographical space in regions that can be further decomposed into subregions. The resulting region hierarchy forms a tree data structure. The most popular indexing methods are Grid, R-tree, and their variants \cite{MicrosoftSQL}\cite{guttman1984r}. 
	
	Grid methods decompose the space into a uniform grid; a tile of the grid can be split iteratively into a given number of smaller tiles, realizing a multilayer hierarchical grid structure. A reference of a spatial object is inserted in the smallest tile that fully contains it. Spatial queries can be carried out, intersecting the requested area with grid tiles and then accessing the contained object references.
	
	R-trees recursively group spatial objects in their minimum bounding rectangle (MBR), thus decomposing the space in a tree of overlapping rectangles, whose number, size, and position depend on the stored spatial objects; leaf nodes contain only one spatial object \cite{guttman1984r}. A range query is carried out using a recursive algorithm; starting from the root node, it goes down in the tree.
	
	The R-tree method is more efficient in terms of storage consumed by the index structure, but the index may change as a function of the inserted items. Conversely, a grid-based index has the advantage that the structure of the index can be created first and data will be added on an ongoing basis without requiring any change to the index structure; however, some nodes of the index can be unused.
	An index is usually built by a centralized algorithm, but being a complex task, literature works also proposed algorithms to build the index using parallel computing approaches (e.g., MapReduce) \cite{JI2014172}.
	
	In the case of federated databases, usually, there are two indexing levels: local and global. The local index is the indexing function embedded in the local DBS. The global index is instead a specific federation functionality used to understand which sites have to be contacted to solve a federated query. 
	
	Recently, the topic of spatial global indexing is gaining research interest for distributed IoT systems since most IoT data are indeed georeferenced.
	In \cite{wang2015experimental}, the authors cope with the problem of IoT sensor/service discovery by creating a spatial index. Each sensor is characterized by the minimal bounding rectangle of the spatial area it covers. Each sensor is also associated with a gateway, and the gateway is characterized by an Approximated Convex Polygon (ACP), which contains all the sensors' bounding rectangles. Every ACP is collected in a global/centralized R-tree structure that is used to support query routing toward gateways. The approach of forming the global index with coarse areas (the ACPs) rather than the whole set of sensor bounding rectangles reduces the index size and the update frequency. Similarly, in \cite{fathy2017distributed}, the authors use a quadtree structure index based on geohash codes, which stores the coverage areas of sensors (no gateway in this case), even though with the coarse resolution related to the configured geohash depth.
	
	We embraced the idea of having a global index as in \cite{wang2015experimental} \cite{fathy2017distributed}. However, in a federated database, we can't assume a spatial data locality. Indeed, every database can contain data located in every part of the world. For this reason, the use of ACP/MBR-related approaches \cite{wang2015experimental} for database sites is not feasible since an MBR/ACP risks to have a huge spatial coverage, thus making useless the global index.
	
	In \cite{iyer2017scalable}, the authors propose a new spatial indexing approach which minimizes index rebalancing while being efficient also in case of skew data, that is, data items that are not uniformly distributed over the space. The proposed solution, based on a multidimensional tree of MBRs, can be distributed over the nodes of a cluster by partitioning the space covered by the nodes of the tree among the different nodes of the database cluster. Although efficient for a cluster/cloud deployment in which RTT are negligible, this algorithm may provide very prohibitive access delay if the nodes of the cluster are distant each other, like it is in a federated environment.
	
	Finally, we point out that ETSI Context Information Management Group is proposing the new IoT NGSI-LD standard \cite{NGSI-LD}, which foresees the federation of context brokers (i.e., servers exposing IoT data) through the use of a centralized global index function named context registry. 
	
	In this paper, we propose a distributed global index strategy based on a constrained and adaptive grid scheme for which every database exposes to others the set of \textit{active} tiles where it has data. This meta information forms the global index, which is replicated over all sites of the federation, thus reducing access latency. Synchronization overhead is limited because only those small sets of object insertions that create new active tiles are followed by a synchronization procedure. Moreover, ICN multicasting and caching make the procedure efficient in terms of networking overhead.

	\subsection {Federated database systems} 
	Even though the problem of database federation has been known since the early '80s, there is not much literature on the topic because in recent years, the research focus has been on different topics, including the "integration" of heterogeneous data/databases (see, for example, AWS Data Lake) into a single, uniform, non-redundant data store; database virtualization and hybridization, that is, the inclusion of SQL and NoSQL components in a single logical database \cite{hybriddb}; etc. 
	
	However, ongoing IoT standardization actions (e.g., NGSI-LD), as well as possible privacy-related regulation issues (e.g., GDPR) preventing to move everything to "sub-processor" premises (e.g., a cloud provider), may revamp the research interest on federated databases since they could have many real applications in the near future. 
	
	The paper \cite{mcleod1980federated} is one of the first works focusing on the database federation concept. The authors discuss, even though not entering in technical details, different design alternatives and a generic sequence of operations that should be performed by a "federal controller" to execute a \textit{transaction} (e.g., a query) through the federation. The controller identifies which are the target DBSs that should satisfy the transaction, translates the request in their format, and then collects the results. We propose a more distributed design and the use of a \textit{neutral} transaction format between the controller and the remote DBSs. In so doing, the introduction of new technology does not require changing existing sites\footnote{We note that because of the complexity of the issue, the design of a neutral format solving the heterogeneity problem mentioned in the introduction is outside the scope of this paper. However, the architecture and networking solutions we are proposing are agnostic to the eventual neutral format.}. Moreover, we also provide an implementation of the proposed concepts, exploiting ICN and considering the issue of global indexing.

	In \cite{laurini1998spatial}, the author proposes solutions for problems arising from the integration of heterogeneous "relational" databases with different schemata and spatial object representations (e.g., different projection schemes, geometric discrepancies for boundary objects, etc.). Moreover, the paper explores two possible solutions for global indexing based on Peano keys and R-trees. Our work follows the position of \cite{laurini1998spatial} regarding the need for a global index, but we use an adaptive grid structure since we believe it is more stable for database operations. 
	We do not consider the problem of heterogeneity of the spatial representations because we assume GeoJSON as the common format; otherwise, the solutions proposed in \cite{laurini1998spatial} can also be used in our framework. Finally, our work is much more focused on the networking aspects and includes a performance evaluation. 
	
	In \cite{dharmasiri2013federated}, the authors presented an architecture for NoSQL databases that is successfully tested with CouchDB, MongoDB, and Cassandra. Different from ours, their system uses classic TCP/IP and query flooding (no global index).

	Finally, we point out that in \cite{detti2017application}, we dealt with ICN and NoSQL databases. However, in that paper, we explored how ICN can be internally used to implement a \textit{distributed} database such as that shown in the right part of Fig. \ref{f:fdbms} (DBS \#n). Instead, in this paper, we explore how an ICN can be used for a different and higher-layer problem: the federation of heterogeneous DBSs. In \cite{detti2017application}, the data items are distributed over the available databases of the cluster according to a sharding logic. The sharding logic divides the storage space into geographical zones and assigns each zone to a single database of the cluster, which will be responsible for storing objects intersecting its zones . In this paper, it is the user who chooses in which database to store her data rather than a sharding logic. Consequently, every database can store every spatial object independently of the object location. This additional degree of freedom, typical of a federated database, is required to change the query resolution process. Indeed, whereas in \cite{detti2017application} the query routing is driven by the sharding logic, in this paper, the query resolution is driven by a new global indexing function that identifies the database to be involved in the query and that has been properly designed and optimized to exploit ICN functionality.
	
	Besides, we note that the ICN distributed database that we proposed in \cite{detti2017application} can be considered as one of the possible distributed DBSs that can join our proposed federated architecture like MongoDB, CouchBase, etc. \begin{figure}[t!]
		\centering
		\includegraphics[scale=0.45]{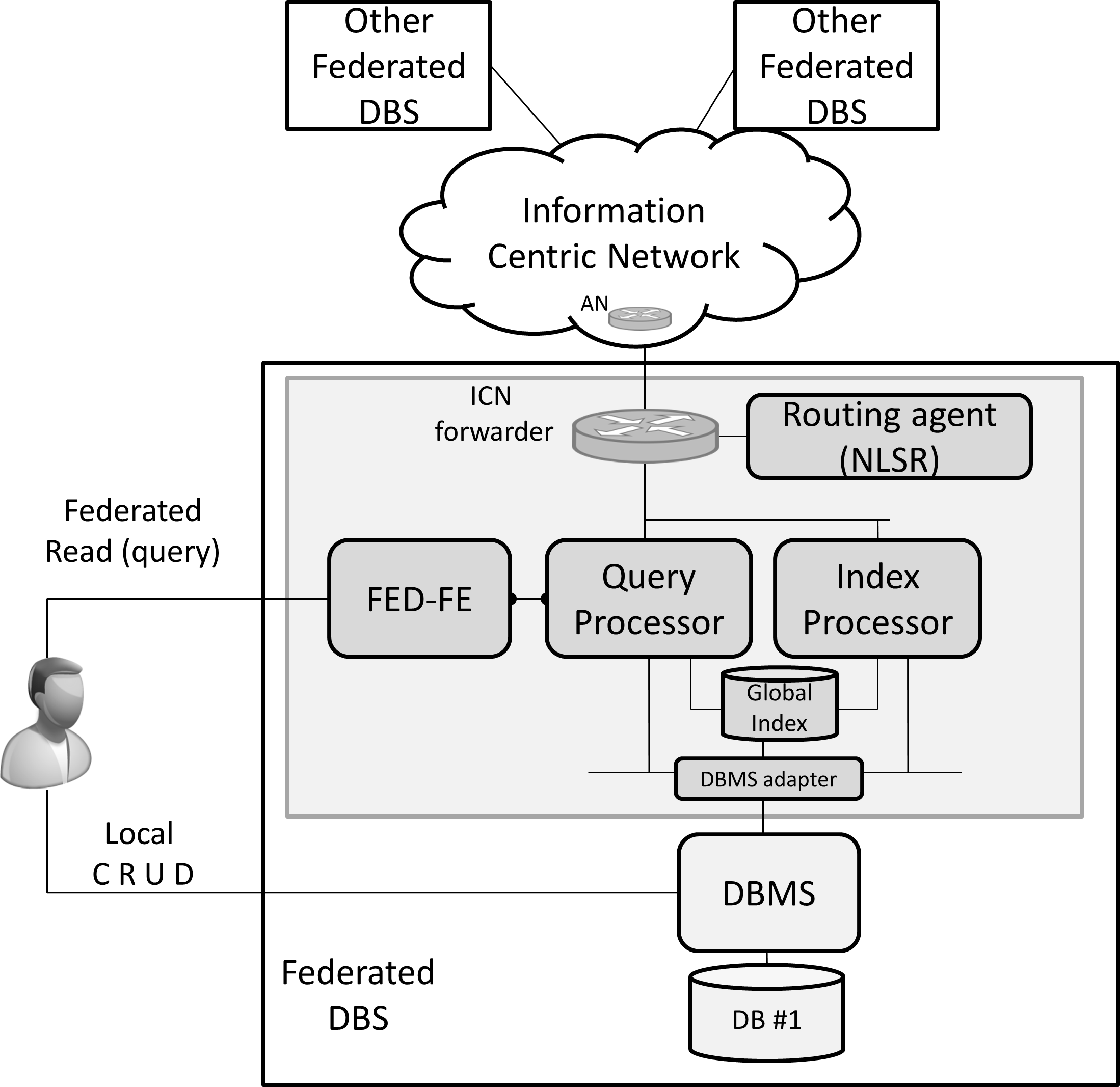}
		\caption{ICN federated database architecture}
		\label{f:arch}
	\end{figure}
	
	\section{ICN federated databases}
	
	We consider a federation formed by autonomous DBSs, each one with its administrator and users, able to store GeoJSON spatial data and to support a common set of spatial operations, such as range queries. Every user is registered to a \textit{home} DBS and can execute Create, Read, Update, and Delete (CRUD) operations only for objects stored in its home DBS and through the local DBMS API (Fig. \ref{f:arch}). The user can access the federation functionality to extend the scope of Read (queries) operations to the overall federation; a federated query searches for matching data through the whole set of federated DBSs as if they were a single one.
	
	Different from SQL ones, NoSQL databases do not (still) have a standard language for querying the DB, and this may potentially hamper a complete NoSQL/SQL federation because only a subset of functionality can be available in some federation sites. In \cite{dharmasiri2013federated}, the authors observe that traditional query operations are supported by most NoSQL databases (as well as SQL ones) and it is possible to translate these operations from one language to another. Concerning spatial databases, we verified that most databases (MongoDB, CouchBase, CouchDB, PostGIS, etc.) support \textit{spatial range queries}, which are searches of objects intersecting or contained in a given geographical area and having specific properties. For instance, a spatial range query may be "search all objects included in the polygon P1 and whose property 'POI type' is equal to hotel." In this paper, we assume that the types of queries that a user can submit to the federation are spatial range queries only because of their wide availability.
	
	The storage space of the federation is organized in data-sets, each of which is identified by unique data-set identifiers (\texttt{did}). A federated data-set is actually the union of the homologous data-sets available in the federated DBSs. For instance, assuming that the federation is formed by two DBSs, both having the data-set "POI," the union of these local data-sets forms the federated data-set with \texttt{did=POI}.
	
	A data-set contains spatial objects structured as generic GeoJSON objects. As made by many databases (e.g., MongoDB), during an object insertion, the system automatically adds an internal property called object name (\texttt{oName}), which uniquely identifies the object in the database system. In our case, the object name does something more because it identifies a specific \textit{version} of the object; thus, if the object content is updated, its object name is automatically changed.
	
	\subsection{Functional architecture}
	Figure \ref{f:arch} shows the functional architecture of the federated database. The large gray box contains the federation functions that we will discuss.
	
	Federated queries are received by a federation front end function (FED-FE) that controls the access rights of the user, executes the requested queries interacting with the query processor, and sends back the answer to the user. 
	The query processor carries out the query by interacting with local and remote DBSs and collecting their answers. Preliminarily, the query processor uses a global index function to single out the subset of DBSs that might have matching objects, thus reducing the query distribution scope. 
	A DBMS adapter is used to "translate" the queries generated by the federation functions (query processor, global index, etc.) in the final language used by the local DBMS.
	
	The communications among federated DBSs are supported by an information-centric network connected to the local DBS through an ICN forwarder. The ICN could be public or private, and it is offered by an ICN service provider. 
	The FIB of the ICN forwarder is automatically configured by an ICN routing agent (e.g., NLSR \cite{hoque2013nisr}) that has a peering relationship with the ICN access node (AN) of the provider.
	
	\begin{table}[]
		\centering
		\caption{Abbreviations}
		\label{t:terms}
		\begin{tabular}{p{1.5cm}p{6cm}}
			\toprule
			Abbreviation & Description \\ 
			\midrule
			DBS & DataBase System that refers to a site of the federation \\
			DBMS & Database Management System that refers to the specific database technology used in a site (e.g., MongoDB, CouchBase, etc.) \\ 
			oName & Unique identifier of an object in the federation change when the object changes \\
			oInterest & Interest packet used to fetch an object through an oName \\
			oData & Data packet used to satisfy an oInterest and containing the requested object \\
			qName & One-time ICN name used to encode a query statement for a remote DBS \\
			qInterest & Interest packet used to send a query statement to a remote DBS using a qName \\ 
			qData & Data packet used to satisfy a qInterest and contain the oNames of objects matching the query statement\\
			vInterest & Interest packet used by a DBS to advertise a new version of the global index information; it is distributed over the multicast prefix \texttt{index/notify} \\
			gInterest & Interest packet used to fetch global index information from a remote DBS \\ 
			gData & Data packet used to satisfy a gInterest and contain global index information of a DBS \\
			dbsid & unique identifier of a DBS of the federation \\
			did & data-set identifier \\
			\bottomrule
		\end{tabular}
	\end{table}
	
	\subsection{Joining procedure}
	To join a new DBS to the federation, the DBS administrator obtains from the ICN service provider a unique database identifier (\texttt{dbsid}) and a valid ICN certificate signed by the provider. The (unsigned) certificate of the provider is the security trust anchor of the federation.
	
	The ICN routing agent (Fig. \ref{f:arch}) establishes a peering relationship with the ICN access node (AN) and announces the prefix \texttt{dbsid}. The routing agent concurrently receives the identifiers of other DBSs of the federation. To support global indexing functionality (discussed later on), the routing agent also announces the prefix \texttt{\{dbsid\}/index/data} and the multicast prefix \texttt{index/notify}. After the propagation of these routing announcements, any Interest whose name contains one of these prefixes will reach the joined DBS. 
	
	To avoid the joining of unauthorized DBSs or the tampering of ICN routing plane information, any routing announcement is signed by the DBS, and the signature is verified at the receiving side, that is, by any other routing agents both in the ICN network and in the remote DBSs.

	\subsection{Query resolution}
	
	\begin{figure*}[t]
		\centering
		\includegraphics[scale=0.3]{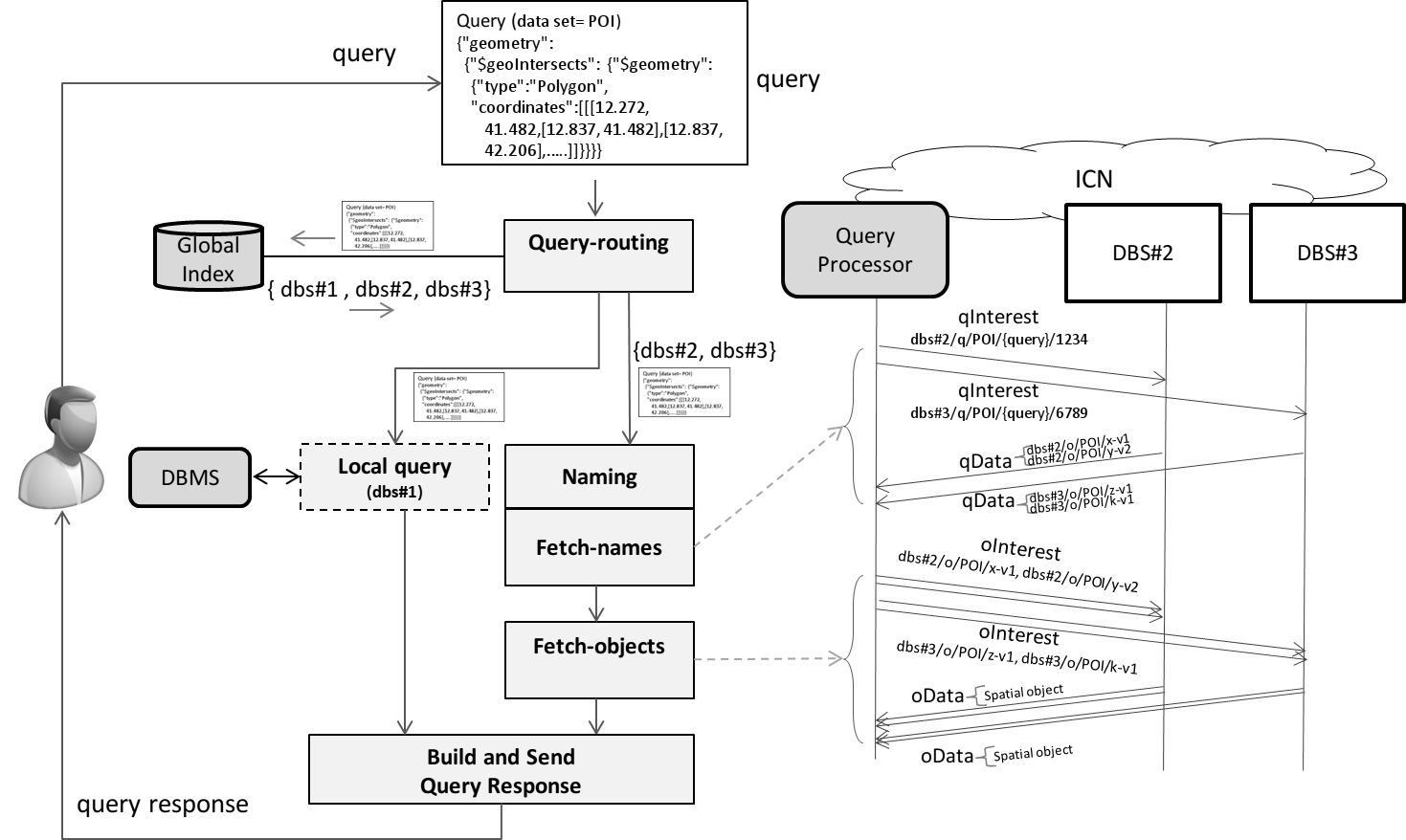}
		\caption{Query procedure executed by the query processor}
		\label{f:qp}
		\vspace{-5pt}
	\end{figure*}
	
	A spatial range query is solved by a query processor as shown in Fig. \ref{f:qp}. In the following explanation, we denote as \textit{local} query processor the query processor receiving the query from the user and \textit{remote} query processors the query processors of remote DBSs participating to the query resolution.
	
	Initially, the local query processor selects the DBSs of the federation deemed worthy to send the query. A simple choice, named \textit{query flooding}, would be to send the query to all the DBSs. Query flooding has the advantage of not requiring any knowledge about what is stored in the databases but has the cons of possibly overloading them with queries for which they haven't matched data. When the federation grows, it is necessary to be more efficient by adopting \textit{query routing} strategies that reduce the query scope by identifying the subset of databases that might have matching objects. To this aim, the local query processor asks the global index function by passing to it the spatial area queried by the user; in turn, the global index function replies with the identifiers of the DBSs having objects in that area (e.g., local dbs\#1, dbs\#2, and dbs\#3 in Fig. \ref{f:qp}). This knowledge comes out from a sharing of metadata among the databases of the federation, for which every database provides to the other a coarse vision of spatial regions (active tiles) in which it has stored objects, as described in Section \ref{s:index}. 
	The local query processor subsequently relays the query to the selected DBSs and eventually collects the query results using the following ICN procedure. 
	
	For relaying a query to a remote DBS, a naming function computes an ICN name, called \textit{qName}, composed of the identifier of the remote DBS (\texttt{dbsid}), the \texttt{q} marker identifying that it is a query name, the data-set identifier (\texttt{did}), the query statement, and a random nonce; for example, \texttt{dbs\#2/q/POI/\{query\}/1234} is a possible qName. 
	For each remote DBS, a qName is computed, and an Interest packet (\textit{qInterest}) is sent out\footnote{We note that all the used Interest and Data packets are unmodified ICN packets. We merely called them qInterest, qData, oInterest, etc., to better identify them during the explanation.}. ICN forwarders route by name a qInterest to the remote DBSs exploiting the database identifier (\texttt{dbsid}) contained in the qName and their FIB entries.
	
	The qInterest is handled by the remote query processor, which extracts the query statement inside the qName and relays the query to the local DBMS thereafter. This query is made in such a way to retrieve only the names (\textit{oNames}) of the objects matching the query conditions rather than the whole objects' information. The resulting list of oNames is packaged in a Data packet called \textit{qData}, which is sent back to satisfy the qInterest. 
	An oName is formed by the identifier of the database system (\texttt{dbsid}), the \texttt{o} marker identifying that this is an object name, the data-set identifier (\texttt{did}), and a unique string that identifies a specific version of the object; for example, \texttt{dbs\#2/o/POI/\{\_id\}-v1} is a possible oName, where \texttt{\{\_id\}} is a unique identifier of the object in the specific DBS, possibly equal to the ID natively provided by the DBMS, and \texttt{v1} is a version number. 
	
	When all qData packets coming from remote DBSs are received by the local query processor, the latter has the whole list of oNames of federated objects matching the query condition. These objects are then pulled through parallel oInterest–oData packet exchanges.
	
	In so doing, we are solving a query in two phases: a \textit{fetch-names} phase and a \textit{fetch-objects} phase (Fig. \ref{f:qp}). This may sound as a temporal inefficiency, but we have chosen this approach to exploit the ICN in-network caching as hereafter discussed.
	
	\subsection{Exploitation of in-network caching}
	Even though caching can dramatically accelerate query resolution, its usage should be carefully designed in database applications when it is not acceptable to send back stale data. For this reason, we used the two-phase query resolution strategy proposed in \cite{detti2017application} and hereafter more widely discussed.
	
	Before devising the two-phase strategy, we had considered a simpler one-phase strategy made of a qInterest–qData exchange only, for which the returning qData packet merely contained the whole set of matching objects. However, we observed that the result of a query statement may change over time because of object insertions or removals. Consequently, in-network caching could not be used for qData to avoid the risk of sending back stale information. In addition, even assuming to agree to receive stale information, cache hits would happen only for those limited sets of queries which are exactly equal each other, for example, two users searching objects on the same area and with the same property constraint. All these mean that the one-phase query resolution strategy cannot be used when the reception stale data is unacceptable and that caching might not be effective because of the natural heterogeneity of query statements. 
	
	For this reason, we decided to move forward to the two-phases strategy, which implies that qData contains only the names of matching objects. These names can change over time; thus, again, qData packet cannot be cached. However, ICN caches can be fruitfully and safely used for next oData packets: fruitfully, because an oData packet contains a single spatial object whose cached version can be reused also by heterogeneous queries, for example, having a partial overlapping of the queried areas, and safely because there is no risk to send back stale data neither after a spatial object update nor after a removal. In fact, when an object changes, the object name (oNames) is changed as well, and such a new name is sent back in the list inside the qData, making it possible to fetch the updated version of the object in the second phase. Moreover, if the object is deleted, its oName will no longer be included in the qData; thus, the deleted object will not be fetched in the second phase. \footnote{We observe that the two-phase strategy also doesn't assure perfect data consistency when the data are updated/deleted in the period of time between qData and oData receptions. Likely, these are rare events because of the short time period in which they can occur, and if these events are unacceptable, ICN caching must be switched off.}
	
	Finally, to avoid cache poisoning problems \cite{bianchi2013check}, Interest and Data packets are signed by the sender and verified by the receiver to avoid access to the federated data from unauthorized DBSs and to avoid information manipulation.

	\subsubsection{Global indexing}
	\label{s:index}
	The federation uses a global indexing function based on a grid strategy and a network synchronization procedure for which every DBS eventually obtains the same version of the index. We choose a grid methodology because it is more stable and reduces the synchronization effort. 
	
\begin{figure}[t]
		\centering
		\includegraphics[scale=0.75]{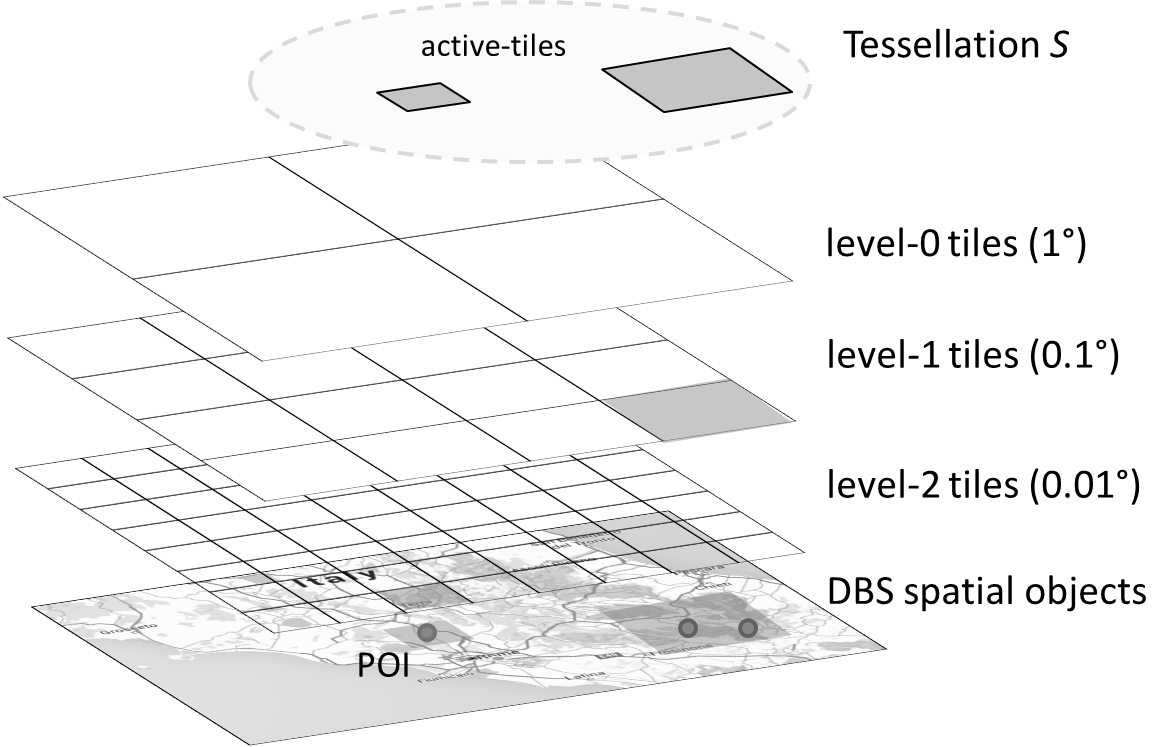}
		\caption{Active tiles of a DBS}
		\label{f:grid}
	\end{figure}

	The grid regions are squared geographical tiles aligned with world parallels and meridians (Fig. \ref{f:grid}). Tiles have $N$ possible resolution levels, from level $0$ to level $N-1$. A tile of level $n$ contains $M$ tiles of level $n+1$. For instance, in case of $M=100$, level $0$ tiles have a lon/lat size of 1 degree, level $1$ tiles have a size of 0.1 degree, level $3$ tiles have a size of 0.01 degree, and so forth. 
	
	The index processor (Fig. \ref{f:arch}) of the $i$-th DBS \textit{tessellates} the area covered by the stored spatial objects with a set $S_i$ of nonoverlapping tiles, denoted as \textit{active} tiles. Thus, a tile is active if it intersects at least a stored object. For instance, in Fig.\ref{f:grid}, we have a DBS storing three POIs. The tessellation $S_i$ has a size $len(S_i)=2$, and it is formed by a level 2 and a level 1 active tile. Because of the insertion, removal, and deletion of spatial objects, the set of active tiles forming the tessellation $S_i$ is periodically updated by the index processor.

	\begin{figure}[t]
		\centering
		\includegraphics[scale=0.7]{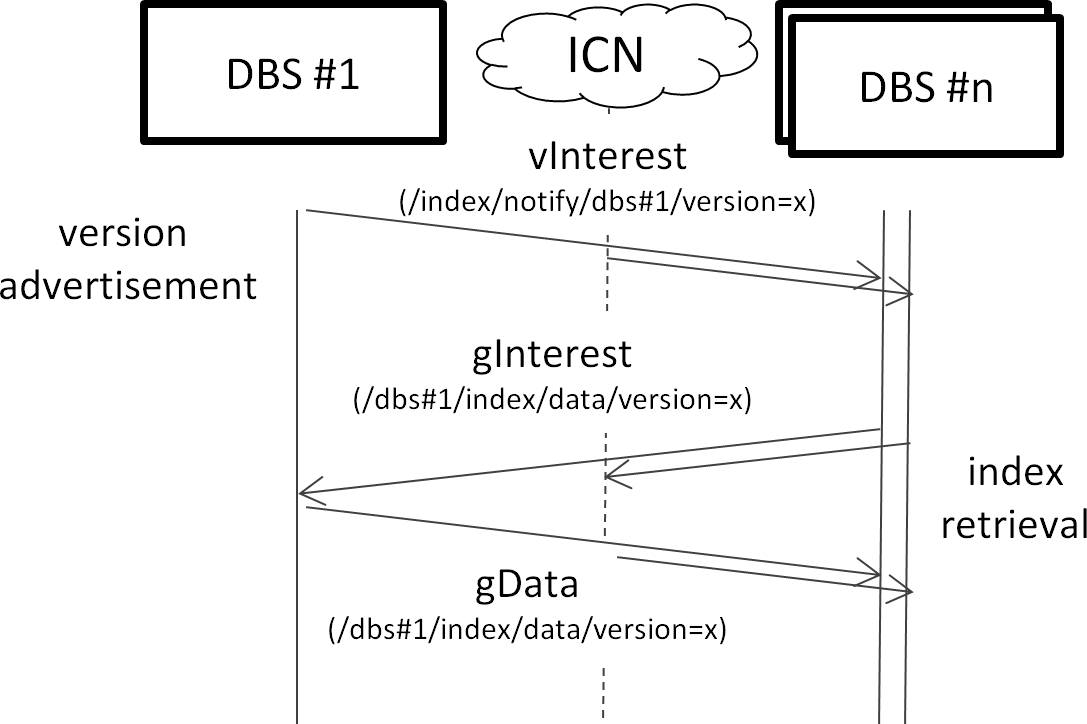}
		\caption{Global index synchronization}
		\label{f:sync}
	\end{figure}

	When the tessellation $S_i$ is updated, the index processor distributes it to other remote DBSs. In parallel, it receives the tessellations $S_j$ of remote DBSs. The global index $G=\bigcup_{i=1} S_i$ is built by each DBS by merging the local and remote tessellations\footnote{Note that different DBSs can have overlapped active tiles just because it is the user who decides where to store her data rather than a sharding logic like in \cite{detti2017application}}. Such a synchronization procedure uses the ICN packet exchanges shown in Fig. \ref{f:sync}, for which DBS \#1 has produced an updated tessellation $S_1$. The procedure exploits ICN multicast and security capabilities as follows: We remind that the routing engine of any DBS announces the multicast prefix \texttt{index/notify} and the unicast prefix \texttt{\{dbsid\}/index/data}. Periodically (e.g., every 1s), the index processor of DBS \#1 sends an Interest packet (\textit{vInterest}) with \texttt{index/notify} as name prefix, followed by \texttt{/dbs\#1/version=$x$}, where $x$ is an increasing version number of the local tessellation $S_1$. The vInterest reaches the ICN network, which, in turn, carries out a multicast distribution toward any other node that announces \texttt{index/notify}, that is, toward any other DBS of the federation. If the receiving DBS \#n has an older version of $S_1$, it fetches the new set by sending an Interest (\textit{gInterest}) for \texttt{dbs\#1/index/data/version=$x$}. The same gInterest can be sent by other DBS nearly at the same time, thus triggering ICN multicast distribution for the returning Data message (\textit{gData}) too, which contains the updated version of $S_1$. To avoid the tampering of the global index or the acquisition of the index from unauthorized entities, any related Interest and Data packets are signed by the producing entity and verified by the receiving one. 
	
	The elements of the global index are stored in a local spatial database, which could be either the same one used for storing spatial objects of the customers or an additional faster in-memory database. Each active tile is stored as a spatial object with a squared shape and with the database identifier (\texttt{dbsid}) as a property. Let us assume that a user submits to the system a range query whose requested area is $A$. To single out which DBSs is to be involved in the query resolution, the query processor submits to the global index function a range query requesting the same area $A$, thereby receiving from the underlying database the set of intersecting active tiles and, in turn, the set of database identifiers of the DBSs to be contacted for solving the query. 
	
	\paragraph{Adaptive tessellation.} 
	Because of the tile heterogeneity, many possible tessellations $S$ may exist, and an optimization problem turns out as follows: 
	When the query processor asks the global index, it obtains a list of candidate DBSs; however, some of them could be \textit{false positives}, which are not actually storing any spatial objects intersecting the queried area. This is because an active tile may be larger than the enclosed spatial objects. For instance, in Fig. \ref{f:grid}, the DBS stores three POIs and advertises two active tiles whose covered area is greater than the POIs'. Consequently, it may happen that a query intersects an active tile but not the contained POI, thus generating a false positive. The consequence of a false positive is the useless sending of queries to DBSs, wasting network bandwidth and, more importantly, processing capacity. The volume of false positives is a measurement of the \textit{accuracy} of the global index: the better the accuracy, the fewer the number of false positives.
	
	A tessellation $S_{min}$ made with the smallest possible tiles, for example, the level 2 tiles in Fig. \ref{f:grid}, has the pros of providing the highest possible accuracy. However, it has the con of generating the highest number of active tiles to synchronize. To give an idea of the involved numbers, if we exclusively use level 2 for a DBS containing 1/3 of OpenStreetMap European POIs, the resulting number of active tiles is in the order of $4 \cdot 10^5$. Such a high number of active tiles has many drawbacks, including a great number of bytes to be transferred during the synchronization process and a higher probability that insertion or removal operations can change the set $S$, therefore increasing synchronization frequency, etc. 
	
	To summarize, smaller tiles provide better accuracy but need a higher synchronization effort. Accordingly, a trade-off shows up, and we can model it with the following optimization problem: finding the best tessellation $S$ with the minimum cost $C_S$, with the constraint that the number of active tiles $len(S)$ is not greater than the given value $k$, that is,
	
	\begin{equation}
		\label{eq:1}
		\begin{aligned}
			& \underset{S}{\text{minimize}}
			& & C_S \\
			& \text{subject to}
			& & len(S) \leq k
		\end{aligned}
	\end{equation}
	
	We define the cost $C_S$ of a tessellation $S$ as the difference between the area covered by its tiles and the area covered by the tiles of the tessellation $S_{min}$ obtained using the smallest possible tiles. The lower the cost, the higher the expected accuracy. Indeed, if we have no constraint on $k$, the best choice (zero cost) in terms of accuracy is to select the tessellation $S_{min}$. We also define the cost $C_t$ of a tile as the difference between the area covered by the tile and the area covered by its children tiles that belong to the minimum tessellation $S_{min}$. It follows that the cost $C_S$ of a tessellation is the sum of the costs $C_t$ of its tiles. In \cite{golab2015size}, the authors demonstrate that such a \textit{size-constrained weighted set cover} problem is NP-hard. Thus, we propose to use the following greedy algorithm \ref{a:ct}.
	
	Initially, the algorithm uses the tessellation $S_{min}$ to build a $N$-level tree $T$, whose nodes are the tiles of $S_{min}$ plus all their parent tiles up to level 0. This tree may have many disjointed roots, and thus, we add a common root node.
	Starting from the tree $T$, the algorithm follows a \textit{top-down} reduction approach, and at the end of the iteration, the active tiles of the tessellation are the leaves of the reduced tree. 
	
	For each step, the algorithm computes which is the highest resolution level of the next tile that has to be added to the final tessellation to respect the constraint. Colloquially, the question posed by the algorithm is "Is it necessary to add a tile of level 0?"; if not, "Is it necessary to add a tile of level 1?"; and so on. 
	In general, it is necessary to add a new level-$i$ tile if we are not able to respect the constraint by finishing the tessellation with all the remaining (smaller) level-$(i+1)$ tiles, but we are able to respect the constraint by finishing the tessellation with all the remaining level-$i$ tiles. 
	
	When the level $i$ of the next tile has been identified, among the tiles of this level not yet included in the tessellation, the algorithm selects the one with the minimum tile cost $C_t$, and the related sub-tree is removed from $T$. Then the iteration restarts and continues until the number of $T$ leaves respects the constraint $k$. 
	There are some exceptional cases in which respecting the constraint is impossible. This happens when even the smallest possible tessellation entirely formed by level 0 tiles has a size $len(S)$ greater than $k$. In these cases, such a minimum size tessellation is returned by the algorithm.

	\begin{algorithm} [t!]
		\caption{Constrained tessellation}
		\label{a:ct}
		\begin{algorithmic}
			\Statex $k$ = max number of active tiles
			\Statex $N$ = number of resolution levels 
			\Statex $T$ = hierarchical tree of tiles
			\Statex $leaves(T)$ = leaves of T
			\Statex $i$ next highest level to be added\\ 
			
			\State Build $T$ from $S_{min}$\\ 
			
			\Statex\textit{\# constraint violation exception}
			\If {num. level 0 tiles of $T > k$}
			
			\Return $S$ = set of level 0 tiles of $T$
			
			\EndIf\\
			\Statex\textit{\# tesselling iteration}
			
			$i=0$
			\While{number of $leaves(T) > k$} 
			\While{$i < (N-1)$}
			\If {(a new level-$i$ active tile is necessary)}
			\State Find the new level-$i$ tile with min. cost $C_t$
			\State Prune all its children from $T$
			\State Break
			\EndIf
			
			$i=i+1$
			\EndWhile
			\EndWhile\\ 
			\Return $S=leaves(T)$ 
		\end{algorithmic}
	\end{algorithm}

	\section{Performance analysis}
	\label{s:perf}
	
	We implemented the architecture in Fig. \ref{f:arch} by using NDN (a specific implementation of ICN) \cite{ndn} and two different NoSQL DBMSs: MongoDB and CouchBase. The related software is available at \cite{github}. The neutral format used to express a federated database operation is JSON, used by MongoDB, properly translated at the receiving side when the local DBMS is different. 
	
	We set up a database federation formed from three sites (i.e., three DBSs) connected to each other by a single ICN node, emulating the network of an ICN provider. We considered two configurations of the federation: in the first configuration, named "1C+2M," a site uses CouchBase, and two sites use MongoDB. In the second configuration, named "2C+1M," we have two CouchBase sites and one MongoDB site. The DBSs and the ICN network node run on different servers connected by a Fast Ethernet switch.

	The data-set stored in the federated DBSs is made up of 3 million European Point of Interests (POIs) gathered from OpenStreetMap. Regarding the distribution of the data items, we considered two types of \textit{locality}: random and country based. In the random case, every POI is stored in a randomly chosen DBS; in case of country-based locality, each DBS stores the POIs of a specific set of countries.
	The countries have been grouped so as to have almost the same number of POIs per DBS. 
	
	In addition to the permanent storage space provided by the databases, that is, MongoDB or CouchBase, data items can be opportunistically stored in the cache of ICN forwarders too, whose capacity is set to 256,000 Data packets, roughly 256,000 POIs. For the global index, we used the three-level hierarchical grid shown in Fig. \ref{f:grid}, and each DBS has a default limit $k$ on the advertised active tiles equal to 20,000.
	
	For the workload, we used \textit{trials} formed by a sequence of 5,000 spatial queries. Each query searches objects located in a squared area randomly centered within European borders. The flow of queries is split among the three DBSs according to a uniform distribution. The query inter-time follows a Poisson distribution, and we define \textit{query rate} the inverse of the average inter-time. Each result is obtained by averaging 10 trials, and all 95\% confidence intervals are smaller than 10\%.

	\begin{figure}[t]
		\centering
		\includegraphics[scale=0.6]{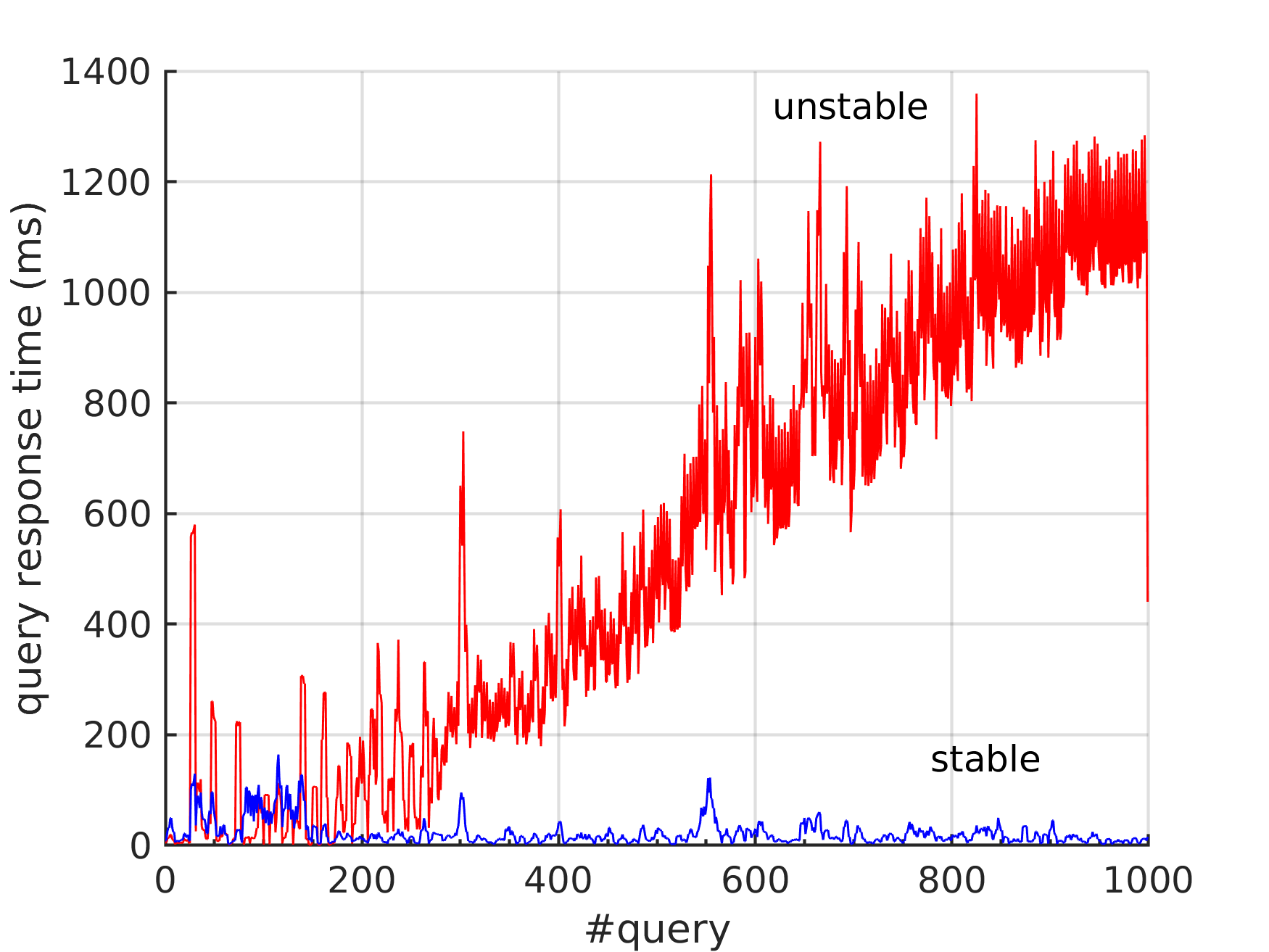}
		\caption{Stable and unstable system behavior}
		\label{f:stab}
	\end{figure}
	\begin{figure}[t]
		\centering
		\includegraphics[scale=0.6]{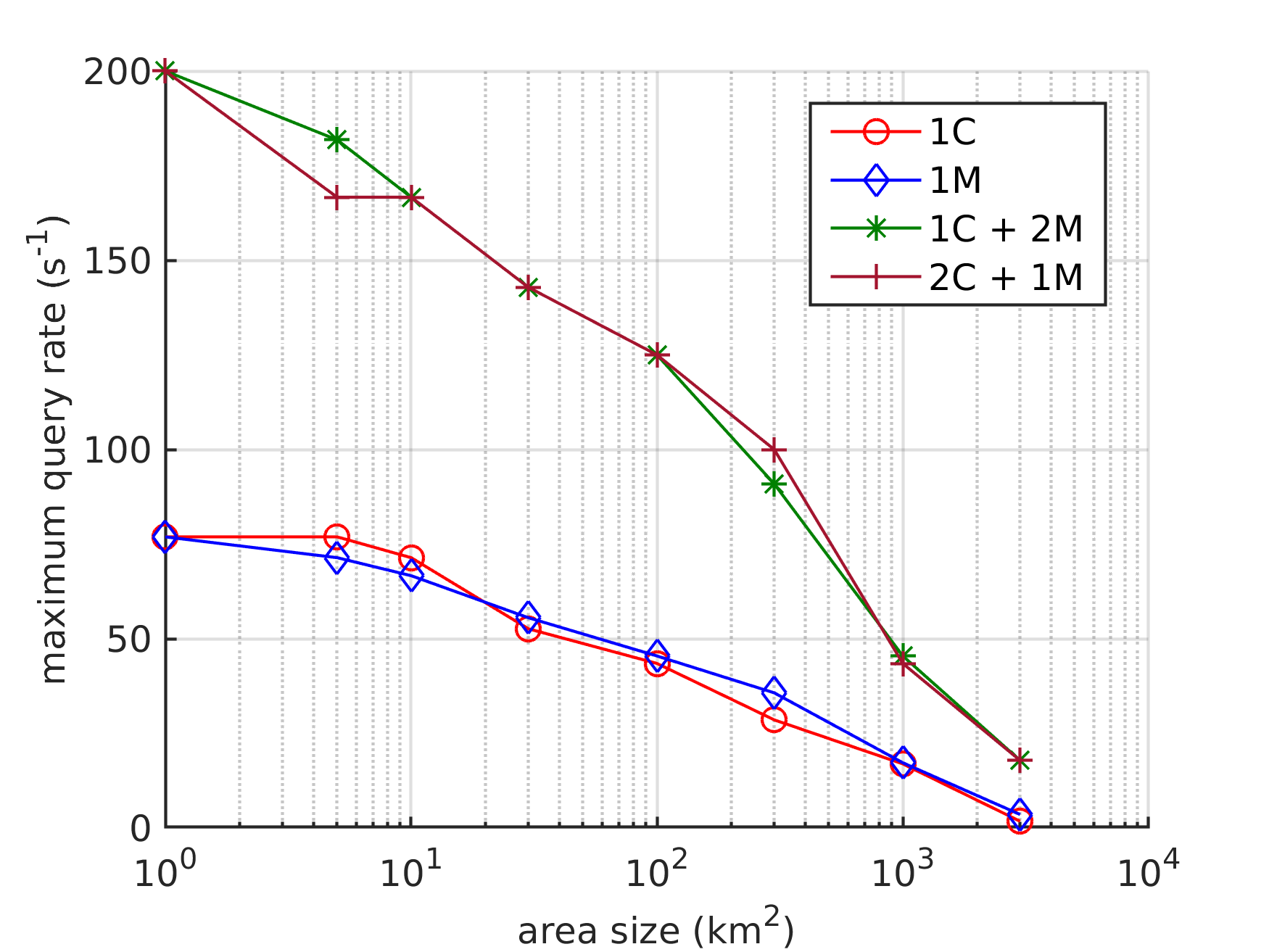}
		\caption{Maximum query rate vs. query area, random storage}
		\label{f:POIscal}
	\end{figure}

	The main performance parameter that we considered is the "maximum query rate" \cite{detti2017application}, which is the highest rate for which the average time needed to solve a query has a stable behavior, that is, the query response time does not have a growing trend versus time. For instance, in Fig. \ref{f:stab}, we report the query response time experienced by the sequence of the first 1,000 queries of a trial. In the unstable case, we used a query rate greater than the maximum one, and we see that the query response time tends to continuously grow like it happens in an overloaded queuing system. Conversely, in the stable case, we used a query rate lower than the maximum one.

	Fig. \ref{f:POIscal} shows the maximum query rate versus the range query area for different configurations of the federation and for a random locality. The rate decreases as the query area increases because the system takes more time to solve larger queries. Indeed, the resolution of larger queries involves a greater number of DBSs, thus increasing their load, and provides query responses with more POIs, thus requiring longer transmission times. 
	
	The two federation configurations, 2C+1M and 1C+2M, verify the ability of the system to smoothly handle heterogeneous DBSs and result in very similar performance. This is because MongoDB and CouchBase sites provide similar performance singularly; thus, a different mix of them does not change the outcome. In fact, the figure shows that a system composed by only one MongoDB DBS (1M) and a system composed by only a CouchBase DBS (1C) provide similar results. We point out that the maximum query rate increases by having more DBSs in the federation (e.g., from 1M to 1C+2M), thereby showing the horizontal scalability of the federated system; the more the available resources are, the higher the sustainable load is.

	\begin{figure}[t]
		\centering
		\includegraphics[scale=0.6]{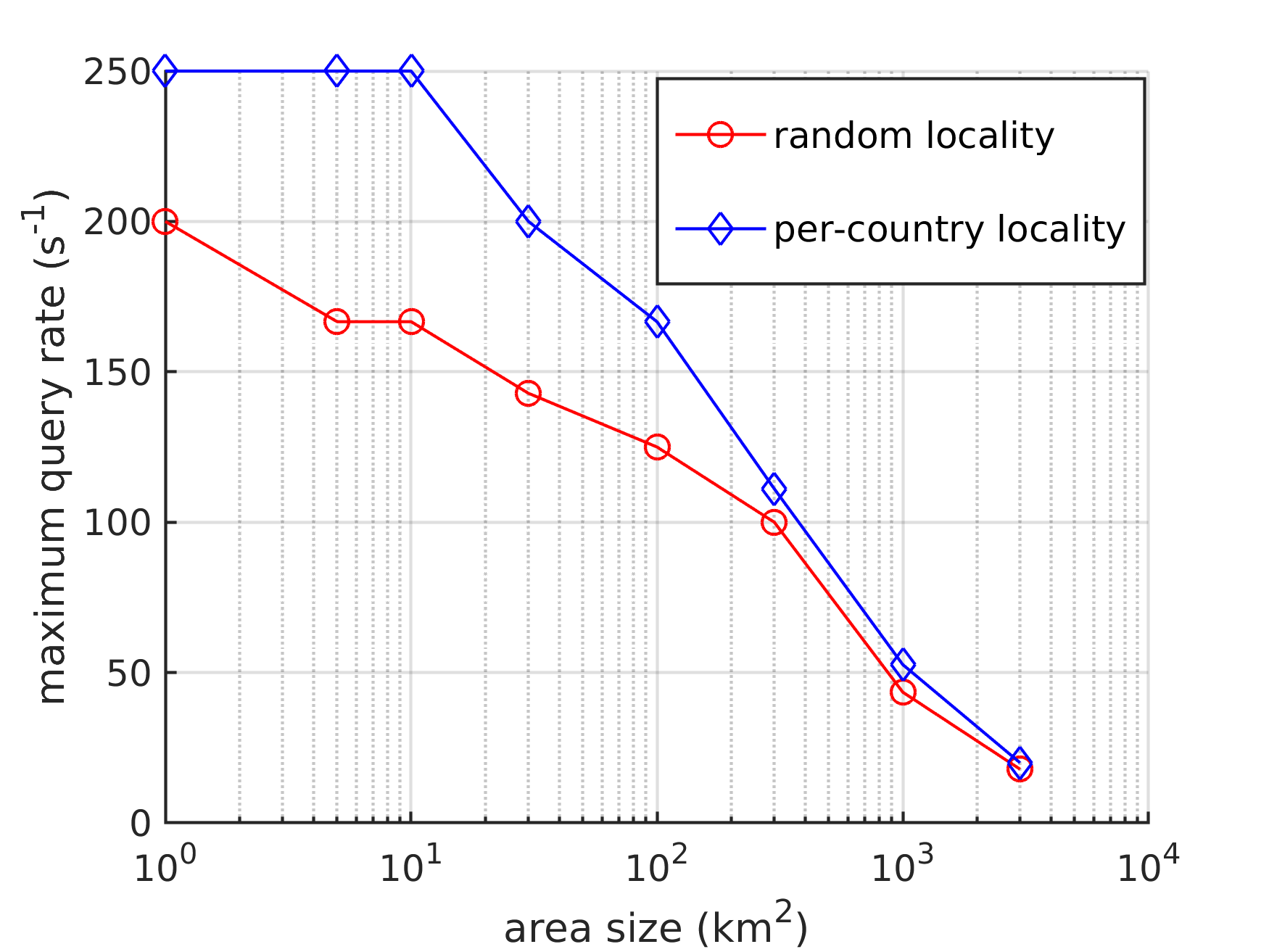}
		\caption{Maximum query rate vs. query area,1C+2M}
		\label{f:POIscalDist}
	\end{figure}
	\begin{figure}[t]
		\centering
		\includegraphics[scale=0.6]{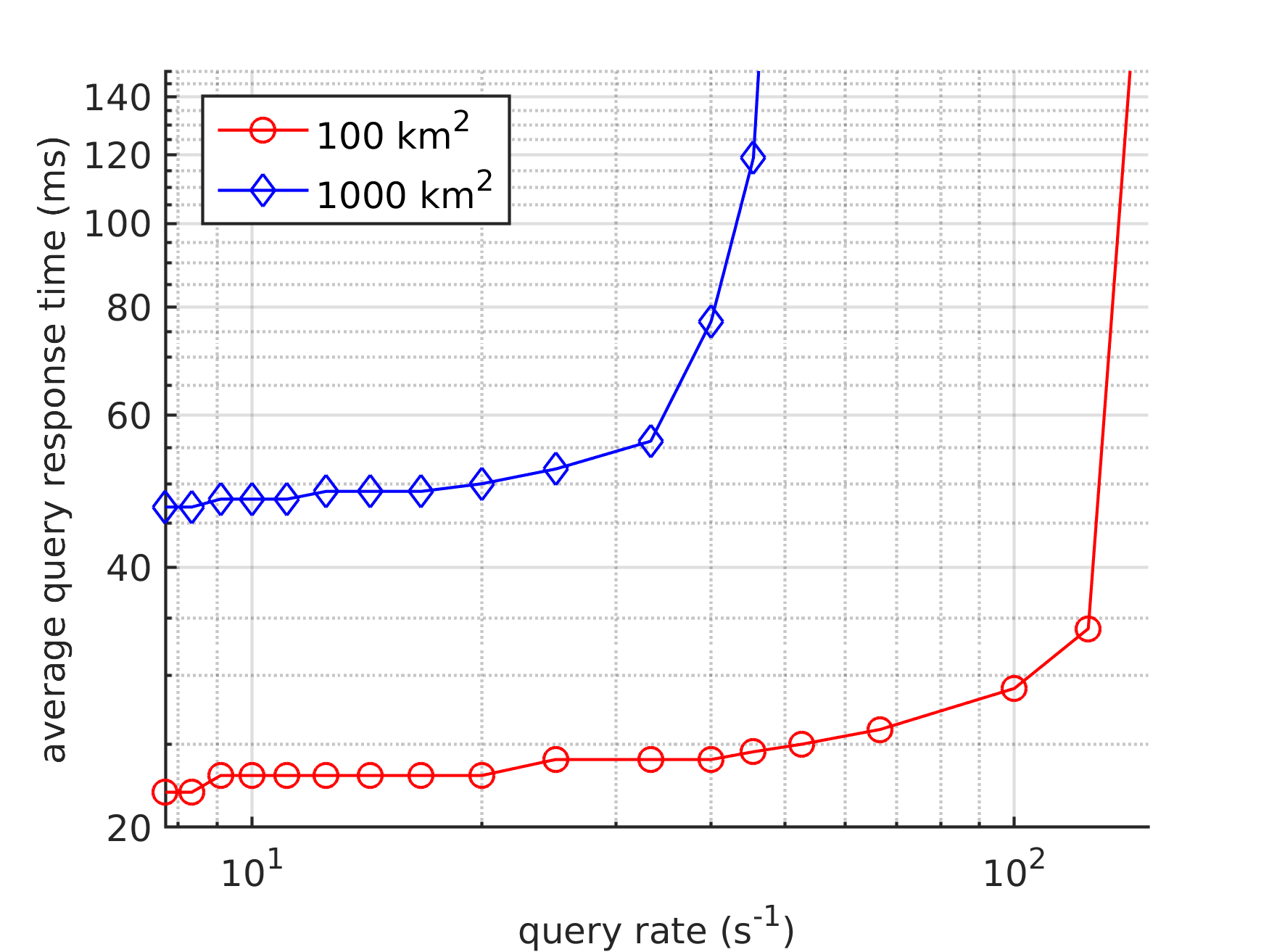}
		\caption{Average query delay vs. query rate, 1C+2M, random storage}
		\label{f:POIqdel}
		\vspace{-0.4cm}
		
	\end{figure}
	
	Fig. \ref{f:POIscalDist} shows the impact of different locality configurations: random and country based. We observe that locality has significant impact on small query areas, up to 100 km$^2$. In these cases, better performance is achieved for a higher data locality (country based) because in this case, the query routing mechanism reduces the number of involved DBSs to solve a query, consequently decreasing the database system's load and response time. This difference fades out as the query area increases because larger queries involve more DBSs; for very large areas, all DBSs are involved in both cases, thus resulting in the same performance.
	
	Fig. \ref{f:POIqdel} shows the average query delay versus the query rate in case of query areas of 100 and 1,000 km$^2$. The query time increases at the increase of the query area and the query rate. The figure also shows that the system provides a stable performance when loaded with a query rate lower than the maximum one, that is, 125 for 100 km$^2$ and 45 for 1,000 km$^2$. When the query rate gets close to the maximum one, the response time quickly grows, and the system becomes unstable \footnote{We note that the current implementation of the query processor uses a thread pool of 4 threads with a waiting queue of 1,000 queries. Accordingly, the system response in Fig. \ref{f:POIqdel} resembles one of a queuing system with multiple servers and a finite waiting queue. Consequently, though not reported in the figure, there is, however, an upper bound for the response time that we measured in the order of 10s.}.

	\begin{figure}[t]
		\centering
		\includegraphics[scale=0.6]{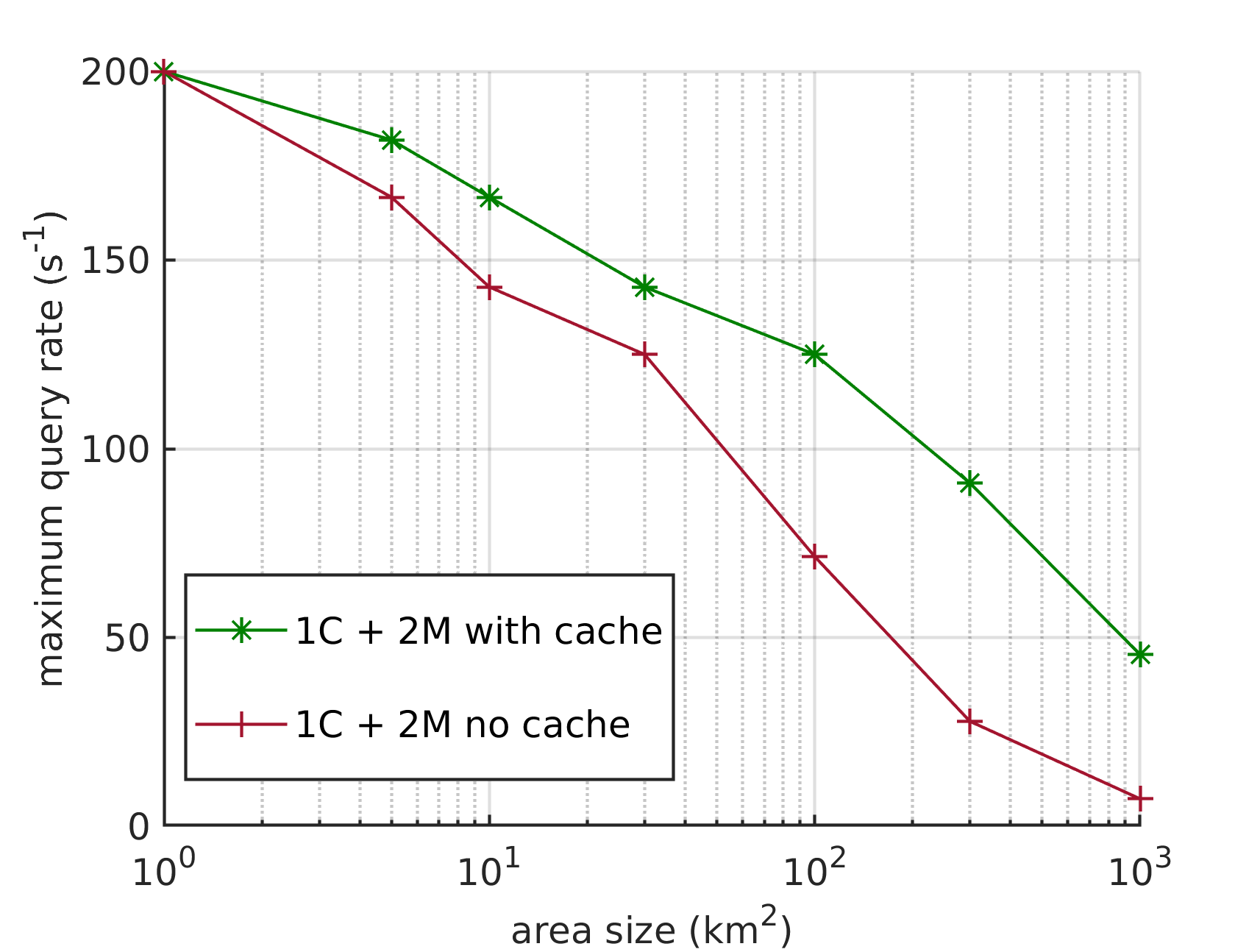}
		\caption{Maximum query rate vs. query area with and without ICN cache,1C+2M, random storage}
		\label{f:caching}
	\end{figure}
	\begin{figure}[t]
		\centering
		\includegraphics[scale=0.6]{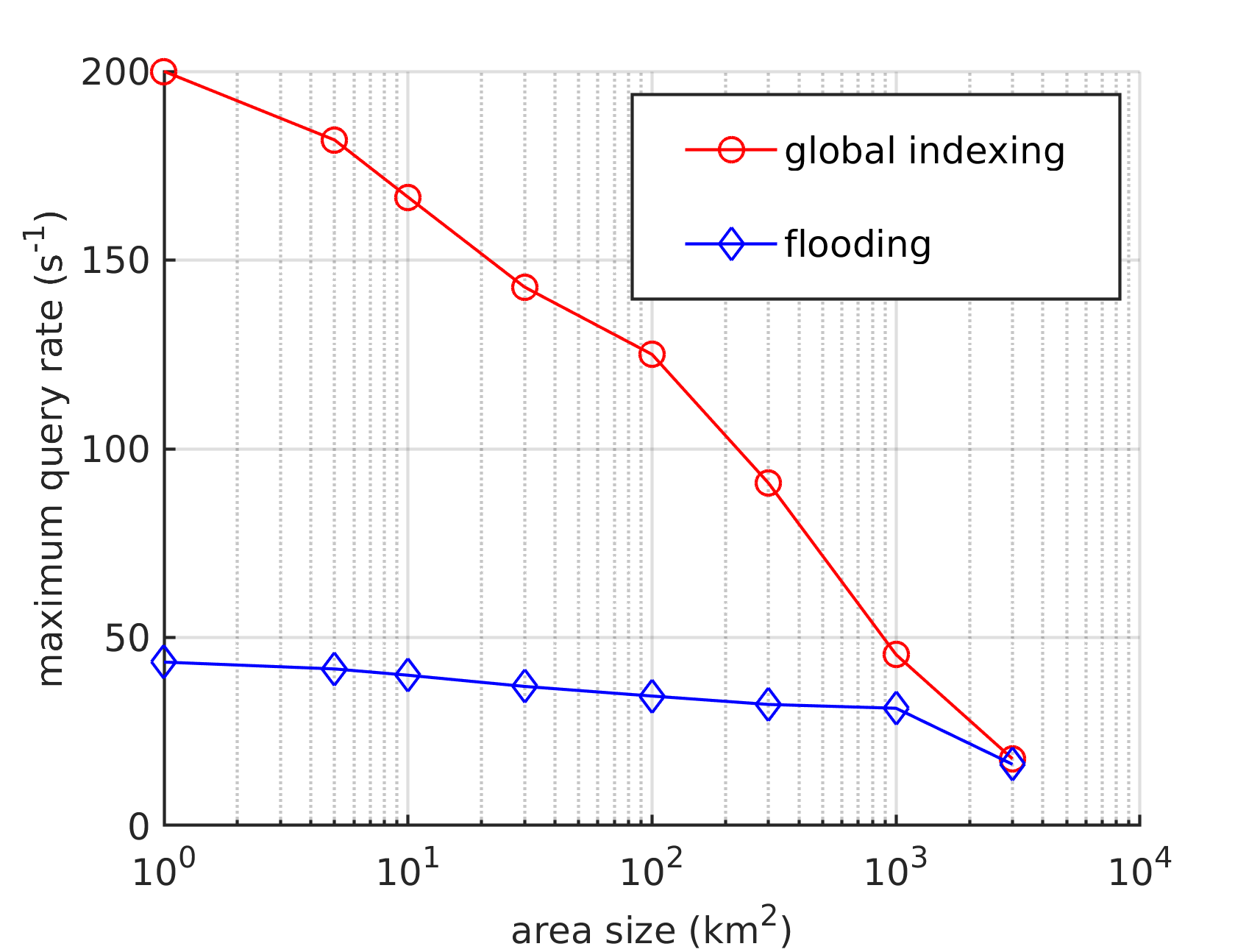}
		\caption{Maximum query rate vs. query area, with and without (flooding) global indexing, random storage, 1C+2M}
		\label{f:tes4}
	\end{figure}
	
	Fig. \ref{f:caching} shows the maximum query rate versus the query area in presence and absence of ICN caching. ICN caches reduce the load of the DBSs and the time needed to fetch matching objects. Consequently, the ICN caching functionality improves the sustainable query rate; this benefit is higher for large queries, which implies the exchange of many objects per query. We note that caching performance depends on the characteristics of the request pattern, including the popularity of objects, the temporal correlation between requests of the same object, etc. \cite{melazzi2014general}\cite{detti2018modeling}. In these measurements, we have considered a flat popularity of the requested objects since range queries are randomly centered in Europe and no temporal correlation since the query inter-arrival times are exponentially distributed. Literature results show that these conditions reduce caching effectiveness, so further improvements are expected for applications characterized by a nonuniform spatial popularity of queried areas.
	
	We now discuss benefits and trade-offs related to the global indexing strategy discussed in Section \ref{s:index}. 
	First of all, we would like to highlight the importance of having a global index, thus justifying the design effort made in this paper. To this end, Fig. \ref{f:tes4} shows the maximum query rate with and without global indexing. In most cases, the resulting rate is dramatically higher with global indexing. Thus, a well-designed global indexing, enabling an effective query routing, can really make a difference in the performance for federated database applications. It is also worth observing that, when the query area becomes very large, the performance with indexing gets close to the flooding one since all DBSs are likely involved, and as a consequence, query routing tends to be useless.

	\begin{figure}[t]
		\centering
		\includegraphics[scale=0.6]{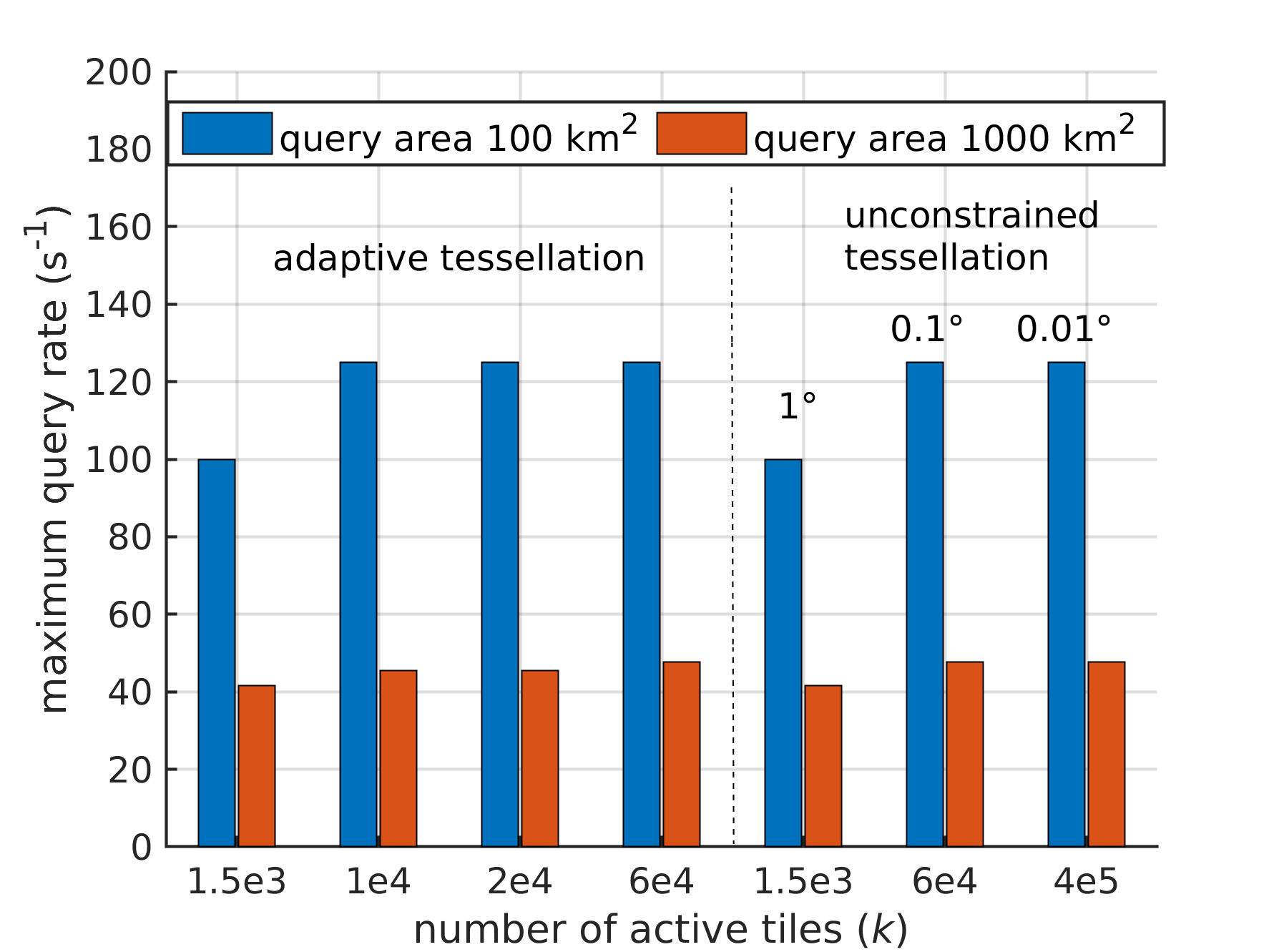}
		\caption{Maximum query rate vs. query area for different values of the number of active tiles and different tessellation, random storage, 1C+2M}
		\label{f:tes1}
	\end{figure}
	
	\begin{figure}[t]
		\centering
		\includegraphics[scale=0.6]{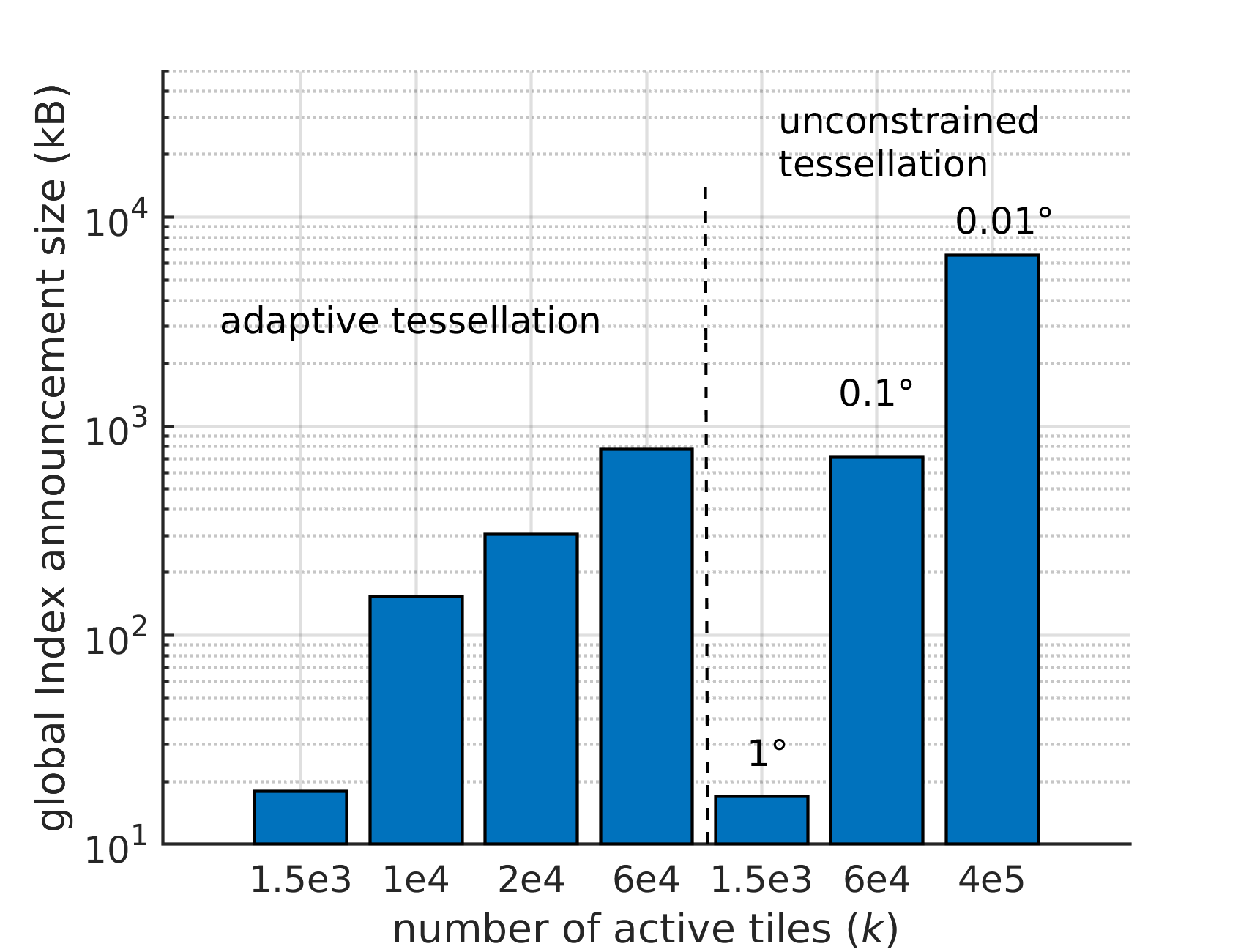}
		\caption{Average size of global index advertisement for different values of the number of active tiles, random storage, 1C+2M}
		\label{f:tes2}
	\end{figure}
	
	Fig. \ref{f:tes1} shows the maximum query rate by varying the constraint $k$ of the adaptive tessellation, from 1,500 tiles up to 60,000 tiles. The figure also includes the results obtained by using a simple unconstrained tessellation having all tiles of the same size for three different configurations of the lat/lon tile size, 1, 0.1, and 0.01 degrees, respectively.
	We observe that a global index with higher accuracy, that is, a greater number of tiles, provides considerable improvements only for small queries (100 km$^2$). This is because there are more false positive events for smaller areas; consequently, a higher accuracy can avoid a significant number of them, thus allowing to support a higher query rate. For larger queries, false positive events become rare; therefore, the impact of higher accuracy is lower. 
	
	By using the unconstrained tessellation, we end up with a global index having about 1,500, 60,000, and 400,000 active tiles per DBSs. These numbers cannot be controlled since they depend on the stored spatial objects. Conversely, the adaptive tessellation allows such control by configuring the parameter $k$. 
	By comparing adaptive and unconstrained tessellation, we see that the former one allows reaching the best performance with 10,000 tiles. Conversely, in the case of unconstrained tessellation, we need about 60,000 tiles to achieve the same result, thus increasing the signaling and processing effort required to maintain the global index.

	A measure of this effort is the average size of the announcements (gData packets) made by DBSs to share their index information; this metrics is shown in Fig. \ref{f:tes2}. As expected, the higher the number of tiles, the greater the announcement size. By comparing the cases of adaptive tessellation with 10,000 tiles and the case of unconstrained tessellation with 60,000 tiles, we observe that the former solution reduces the signaling overhead of roughly 80\% while providing the same performance (Fig. \ref{f:tes1}).
	
	\begin{figure}[t]
		\centering
		\includegraphics[scale=0.6]{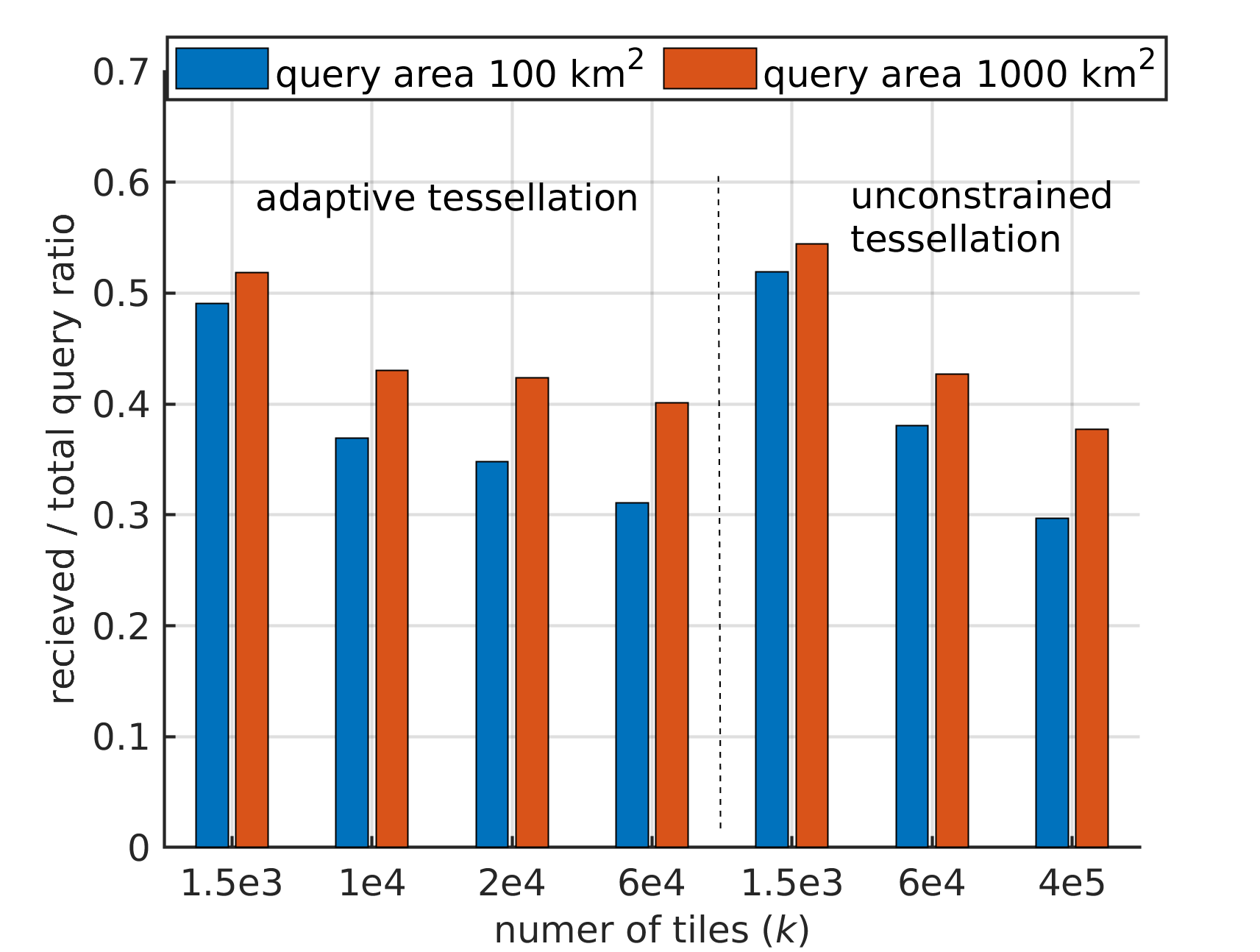}
		\caption{Ratio between the total query submitted to the federation and the query received by a single DBS out of three, random storage, 1C+2M}
		\label{f:tes3}
	\end{figure}
	
	Finally, Fig. \ref{f:tes3} shows the ratio between the total number of queries submitted to the federation and the number of queries subsequently received by a single DBS. If we used query flooding, any query would be sent to any DBS, thus making such ratio equal to one. Therefore, the reported ratio is a measure of the load reduction achieved, thanks to the global indexing and query routing mechanisms. In the worst case, the use of the global index reduces by about 50\% and even less when we increase the accuracy of the index, that is, $k$. The reduction is higher for smaller query areas where the index accuracy causes a greater impact. By comparing adaptive and unconstrained tessellations, we see that the adaptive tessellation provides a lower DBS load for a similar amount of tiles. 
	
	We conclude this section by pointing out that, with the exclusion of Figs. \ref{f:tes2} and \ref{f:tes3}, the obtained numerical results clearly depend on the hardware and on the software code that we have used, being the evaluation based on a real implementation. Consequently, all the above results are of interest more for the insights that they provide than for their absolute numerical values \footnote{After extensive searches in the literature and in the Web, we did not find software solutions for federating heterogeneous NoSQL databases, and for this reason, we have been unable to carry out a comparative analysis.}. 
	
	\section{Conclusions}
	We have shown that multicast, caching, and data-centric security off-the-shelf functionality offered by ICN are useful instruments to cope with communication issues arising from the federation of spatial databases. 
	In fact, the results obtained from a practical implementation and with real data-sets have shown the ability of ICN to effectively federate heterogeneous databases while providing efficient query resolution through global indexing and in-network caching.
	
	It is worth to remind that the federation has some limitations with respect to the use of a single database. First, a federation can fully support only the subset of spatial operations that can be resolved by any SQL/NoSQL participating database. For instance, the NoSQL MongoDB only supports four geo-functions (\$geoIntersects, \$geoWithin, \$near, and \$nearSphere), and the SQL PostGIS supports  more than one thousand geo-functions, including those of MongoDB; thus, federating them limits the exploitable geo-function to the ones of MongoDB. As a second limitation, we observe that the federation introduces latency because of network RTT and processing (e.g., query routing without a preconfigured sharding logic), and it requires synchronization of metadata among databases (e.g., the global index). Consequently, the rate of federated operations is limited too, and for highly intensive database applications, other solutions should be considered. For instance, in the case of rather static data-sets, the integration of all of them in a single big local database will likely provide better performance, but with the cost (and potential privacy issue) of moving all the data in a single point and maintaining them synchronized.
	
	As future directions of this work, we observe that even though we focused on spatial databases, the proposed ICN federated architecture can be extended also for databases storing plain objects but changing the global indexing strategy. To this end, a possible approach is to generalize the index tessellation process as follows: Let us assume that we want to index a specific key of stored objects, for example, the surname of a person, for an address book application. We can map the index key (surname) to a 1D hash space and then tessellate it with the same proposed adaptive algorithm but using 1D segments rather than 2D tiles. In so doing, we will obtain again the benefit of query routing shown in the spatial database case.

	\section*{Acknowledgment}
	This work is supported in part by H2020 EU-JP Fed4IoT project (www.fed4iot.org, EU contract 814918). The document reflects only the author's view, European Commission and Japanese MIC are not responsible for any use that may be made of the information it contains
	
	\bibliographystyle{IEEEtran}
	\bibliography{paper}

	\EOD
	
\end{document}